\documentclass[journal]{IEEEtran}
\usepackage{graphicx}
\usepackage{CJK}
\usepackage{color}
\usepackage[position=t,singlelinecheck=off]{subfig}
\usepackage{caption}
\usepackage{epsfig}
\usepackage{epstopdf}
\usepackage{amssymb}
\usepackage{amsfonts}
\usepackage{amsmath}
\usepackage{mathrsfs}
\usepackage[numbers,sort&compress]{natbib}
\usepackage{epsf}

\usepackage{amsmath}
\usepackage{booktabs}
\usepackage{multirow}
\usepackage{array}
\usepackage{soul}
\usepackage{xcolor} 

\ifCLASSINFOpdf
\else
\fi
\begin{document}
%
\title{Adaptive Diffusion Processes of Time-Varying Local Information on Networks}
%
%
%

\author{Ruiwu~Niu,~\IEEEmembership{}
        Xiaoqun~Wu,~\IEEEmembership{}
        ~Jun-an~Lu,~\IEEEmembership{}
		and~Jinhu~L\"{u},~\IEEEmembership{Fellow, IEEE}~
\thanks{R. Niu, X. Wu, and J. Lu are with the School of Mathematics and Statistics,
Wuhan University, Hubei 430072, China. (e-mail: xqwu@whu.edu.cn).}
\thanks{J. L\"{u} is with the School of Automation Science and Electrical Engineering and Beijing Advanced Innovation Center for Big Data and Brain Machine Intelligence, Beihang University, Beijing 100083, China, (e-mail: jhlu@iss.ac.cn).}
\thanks{Manuscript received XXXXXXXXXXXXXXXX}}

%
%

\markboth{XXX}%
{Shell \MakeLowercase{\textit{et al.}}: Bare Demo of IEEEtran.cls for IEEE Journals}
%



\maketitle

\begin{abstract}
This paper mainly discusses the diffusion on complex networks with time-varying couplings. We propose a model to describe the adaptive diffusion process of local topological and dynamical information, and find that the Barabasi-Albert scale-free network (BA network) is beneficial to the diffusion and leads nodes to arrive at a larger state value than other networks do. The ability of diffusion for a node is related to its own degree. Specifically, nodes with smaller degrees are more likely to change their states and reach larger values, while those with larger degrees tend to stick to their original states. We introduce state entropy to analyze the thermodynamic mechanism of the diffusion process, and interestingly find that this kind of diffusion process is a minimization process of state entropy. We use the inequality constrained optimization method to reveal the restriction function of the minimization and find that it has the same form as the Gibbs free energy. The thermodynamical concept allows us to understand dynamical processes on complex networks from a brand-new perspective. The result provides a convenient means of optimizing relevant dynamical processes on practical circuits as well as related complex systems.
\end{abstract}

\begin{IEEEkeywords}
Diffusion, Synchronization, Entropy.
\end{IEEEkeywords}

%
\IEEEpeerreviewmaketitle

\section{Introduction}
%
%
%
%
\IEEEPARstart{I}{n} the past few decades, network theory has been proved to be a useful tool for modeling the structural and dynamical properties of complex systems~\cite{Barabasi1999,WS1998,ER1960}. From a dynamical point of view, many behaviors taking place in biological, social and technological systems can be considered as diffusion processes on networks, such as disease spreading, rumor transmission and information routing~\cite{Newman2010}. Therefore, scientists have studied diffusion processes, such as random walks, epidemic spreading and synchronization, and perfectly predicted dynamical properties of underlying systems~\cite{Masuda2017,Pastorsatorras2015,Arenas2008}. Alternatively, by observing real systems and analysing big data, scientists have obtained a mass of information from real networks like neural networks, Facebook, transportation networks in a city, and so on~\cite{Barabasi2016}.

To model diffusion processes, random walks are frequently used in a variety of realms, and have been found to have interesting relationships with network topologies~\cite{Masuda2017}. Since Karl Pearson firstly came up with this concept~\cite{Pearson1905}, random walks have been a hot spot in the field of complex networks. This model helps scientists uncover various types of structural and dynamical properties of complex networks~\cite{Porter2014}, and gradually becomes one of cores of network theory. In addition, in order to study spreading processes of diseases, Kermack and McKendrick proposed the Kermack-McKendrick model in 1927, and defined the modern mathematical modeling of infectious diseases~\cite{Kermack1927}. Years later, more realistic descriptions of disease spreading named SIS and SIR models were introduced~\cite{Pastorsatorras2001}. Besides, lots of modified disease spreading models based on previous works were used under different circumstances, even on multi-layer networks~\cite{Wei2018}. These models can approximately describe real-world diffusion processes. Another good approximation for diffusion processes is synchronization. Watts and Strogatz studied synchronization no networks and revealed plenty of properties~\cite{WS1998}, and thereby paved the road to network theory for successors. Sparked by these studies as well as many others, researchers have gone further to understand dynamical processes on both single- and multi-layer networks~\cite{Arenas2008,Liyang2015,Tang2016,Mei2017,Zhuang2018}.

In order to understand the intricate diffusion processes on complex networks, scientists often consider the couplings between a pair of nodes as a constant value~\cite{Almendral2007}. However, in real complex systems, the coupling between nodes sometimes varies with some specific factors, such as evolution time, network topologies, nodal dynamics and so on. Therefore, many researchers want to generalize the coupling patterns of diffusion processes on networks and introduce some intriguing models under different scenarios~\cite{Daido1987,Lumer1991,Filatrella2007}. Daido firstly proposed a model in which the couplings among the oscillators are totally random and found that randomness can cause decrease in the order parameter~\cite{Daido1987}. Later on, Lumer et al. studied a branching tree model under the hypotheses that the couplings vary with the distance and are distributed over a hierarchy of values~\cite{Lumer1991}. In 2007, Filatrella et al. introduced a modified Kuramoto model that accounts for change of the coupling with the number of active oscillators and predicted that both first and second order phase transitions are possible, depending upon the parameter that tunes the coupling among the oscillators. Numerous results show that different couplings can cause different phenomena during the same kind of dynamical processes.

In this paper, sparked by Barahona and Pecora's work~\cite{Barahona2002}, we propose a modified diffusion model, in which the couplings between nodes are varying according to local topological and dynamical information. The outline of the paper is as follows. In Section II we present the modified diffusion model and simulate it on different networks and gain some interesting results. In Section III we investigate the influence of topology on the dynamical processes. In order to further understand the dynamics of the model we proposed, we introduce the idea of state entropy in Section IV. This is not the first time for scientists to analyse complex networks with statistical mechanics methods~\cite{Park2004,Dehmer2011}. Back to 2004, Park and Newman introduced a series of concepts in statistical physics for network analysis~\cite{Park2004}, and opened up a new way to study complex networks. In information sciences, various concepts of entropy were proposed~\cite{Dehmer2011}, and has become a useful tool to reveal the topological and dynamical properties of diffusion processes on networks. Interestingly, we study state entropy of networks during diffusion processes and find that the diffusion of our proposed model is a minimization process of state entropy restricted by Gibbs free energy. Section V presents some conclusions.

\section{Adaptive diffusion processes of time-varying local information}
In social media like Facebook, Twitter or Weibo, there always exists hot topics. We call it a hot topic when the topic can rapidly spread throughout a network. Users' involvement in a topic can be used to measure whether a topic is hot or not. Generally speaking, a user's involvement in the topic can be affected by two factors: the internal factor and the external factor. The internal factor is up to the user's inherent interests in a topic, and the external factor depends on the neighbours' involvement in the topic. Accordingly, we propose an adaptive diffusion model of time-varying local information below,

\begin{equation}\label{1}
\frac{dx_{i}}{dt}=D_{i}\sum^{n}_{j=1}a_{ij}(x_{j}-x_{i}),
\end{equation}
where \(a_{ij}\) is the element of the adjacency matrix describing a network consisting of \(n\) nodes. Node \(i\)'s involvement is represented by \(x_{i}\) (\(0\leq x_{i}\leq 1\)), which can be considered as node \(i\)'s state, and its initial state can be considered as the inherent state. The diffusion coefficient is \(D_{i}=k^{\prime}_{i}/k_{i}\), where \(k^{\prime}_{i}=\sum^{n}_{j=1(x_{j}>x_{i})}a_{ij}\) and \(k_{i}=\sum^{n}_{j=1}a_{ij}\). In this model, when every state of node \(i\)'s neighbors is larger than \(i\)'s, the external factor will play a leading role, and the state of the node \(i\) will accordingly increase. On the other hand, when each of node \(i\)'s neighbors has a state smaller than \(i\)'s, the internal factor will play a leading role, and the state of node \(i\) will stick to its original status. When neighbors with larger and smaller states exist simultaneously, the state of node \(i\) may increase or keep unchanged.

In order to measure the global state of a network at time \(t\), we consider the expectation (average) of all nodes' states as a global parameter, as follows,

\begin{equation}\label{2}
R(t)=\frac{\sum^{n}_{i=1}x_{i}(t)}{n}.
\end{equation}

The simulation is obtained over 20 realizations of randomly generated networks, in which the node size \(n=10000\), the average node degree \(\langle k\rangle=4\), and the time step for integrating the ordinary differential equation (1) using the Runge-Kutta method is 0.01.

\begin{figure}[!t]
  \centering
 \includegraphics[width=2.5in]{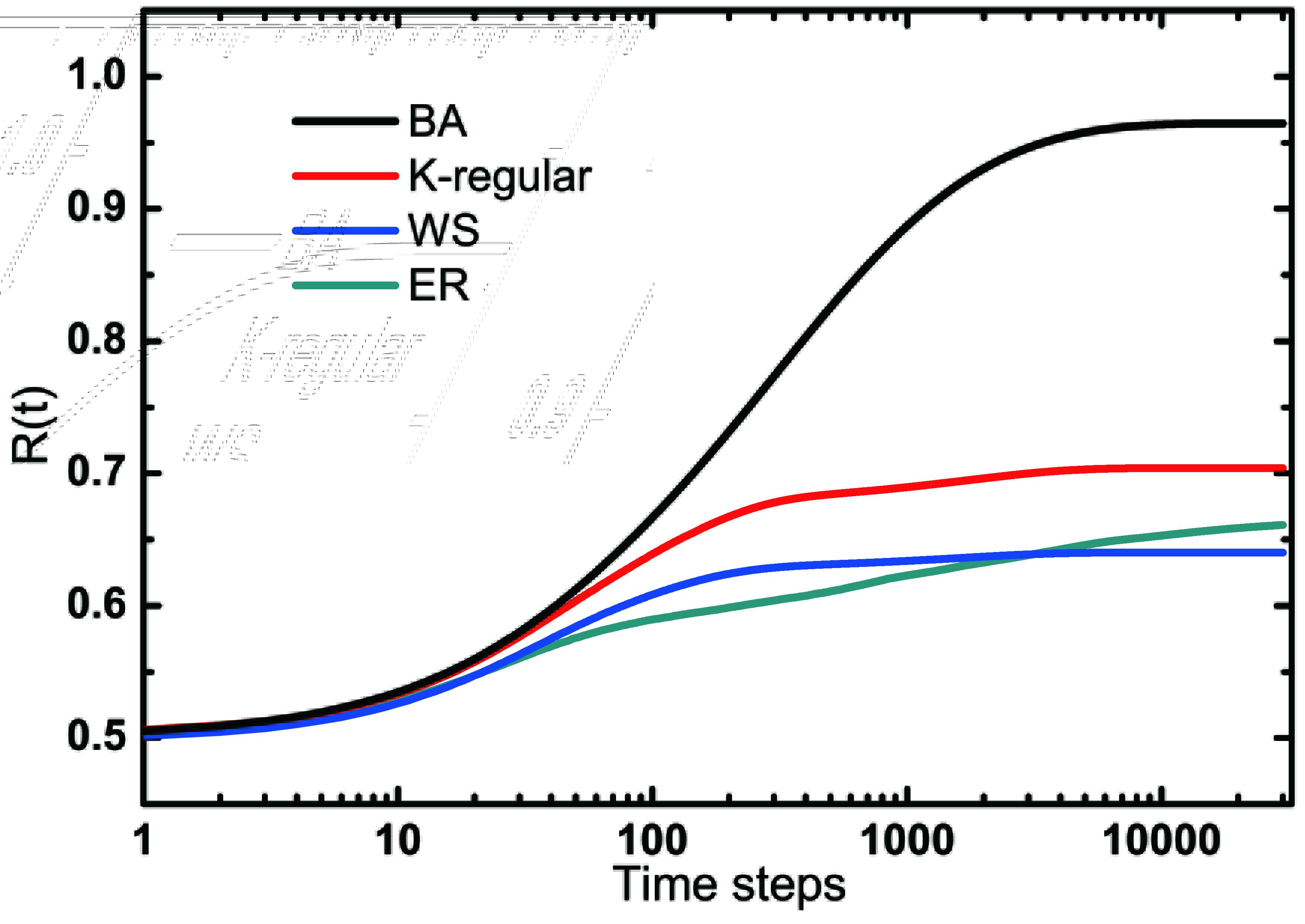}\\
     \caption{The global state parameter \(R(t)\) as a function of time steps.}\label{RT}
\end{figure}

Figure 1 shows the global parameter \(R\) varying with time on different kinds of networks, including the Barabasi-Albert scale-free network (BA network)~\cite{Barabasi1999}, the Watts-Strogatz small-world network (WS network)~\cite{WS1998}, the Erd\"{o}s-R\'{e}nyi random network (ER network)~\cite{ER1960}, and the K-regular network. As the diffusion process evolves, \(R\) increases with time and reach a stable state after some transient time. The difference is, for the BA network, the state of nodes can reach a value that is close to its upper bound \(1\), while for relatively homogeneous networks such as the K-regular, the WS or the ER network, \(R\) can not reach such a large value as the BA network does. We obtain similar observations in the networks with different mean degrees (\(\langle k\rangle=2, 4, 6\)). This means that the structure of the BA network is more beneficial to the diffusion process than the other three models are.

This interesting phenomenon attracts our attention to explore mechanics underlying the diffusion processes. Enlightened by Ref.~\cite{Almendral2007} , we write Eq. (1) into the following matrix form,

\begin{equation}\label{3}
\dot{\mathbf{X}}=-\mathcal{L}\mathbf{X},
\end{equation}
where \(\mathbf{X}=(x_{1},x_{2},\cdots,x_{n})^{T}\) and

\begin{equation}\label{4}
\mathcal{L}=\begin{bmatrix}
D_{1}(\sum_{j}a_{1j}-a_{11}) & \cdots & -D_{1}a_{1n} \\
\vdots & \ddots & \vdots \\
-D_{n}a_{n1} & \cdots & D_{n}(\sum_{j}a_{nj}-a_{nn})
\end{bmatrix}.
\end{equation}

\begin{figure}[!t]
  \captionsetup[subfigure]{labelformat=simple}
  \centering
  \includegraphics[width=2.5in]{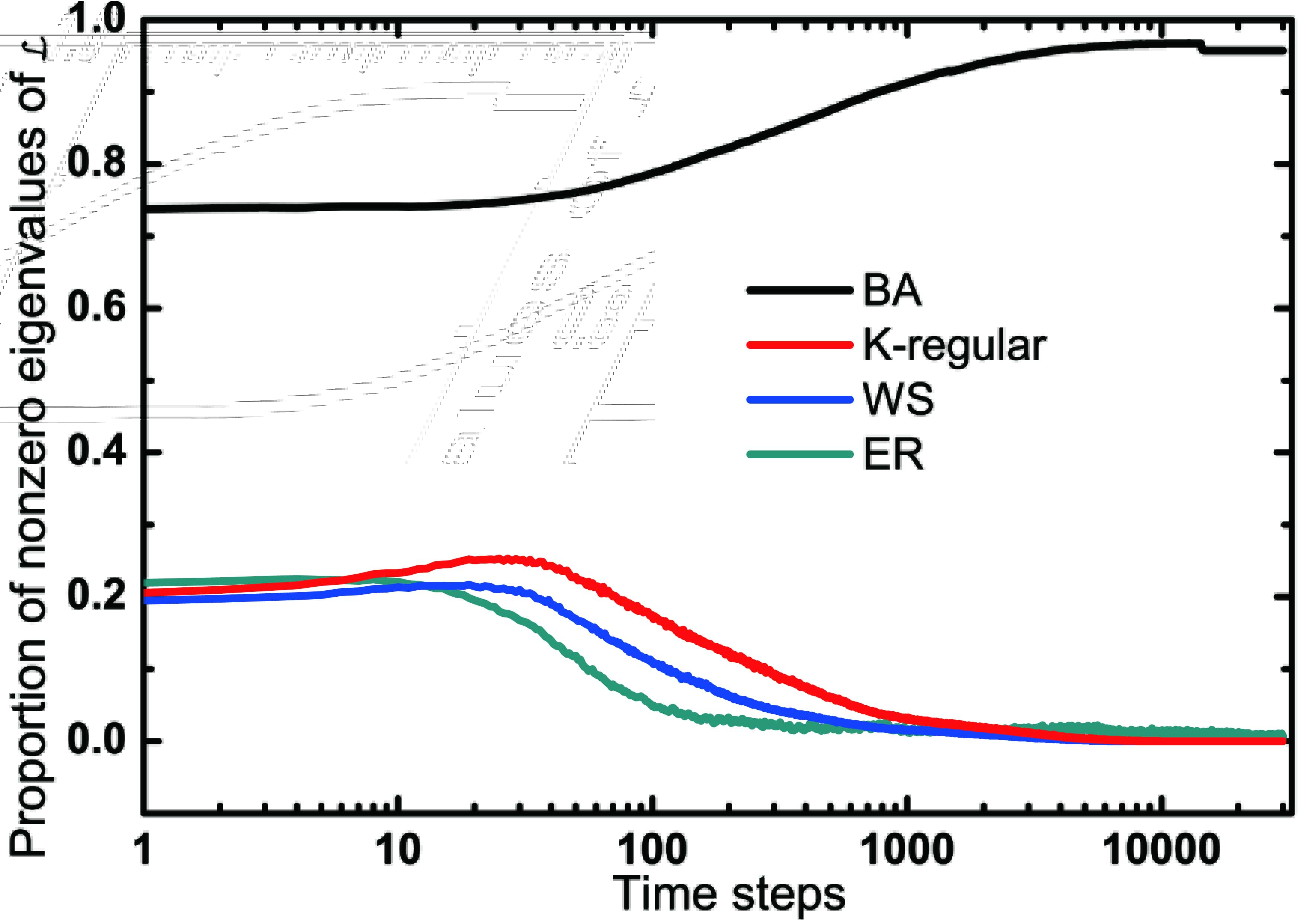}
  \caption{The proportion of none-zero eigenvalues of \(\mathcal{L}\) as a function of time steps \(t\).}
  \label{fig:fig2}
\end{figure}

Different from Ref.~\cite{Almendral2007} , the Laplacian matrix \(\mathcal{L}\) is not necessarily symmetric and changes over time. Then the matrix \(\mathcal{L}\) can be the Laplacian of a directed and weighted time-varying network. The proportion of the none-zero eigenvalues of \(\mathcal{L}\) is also varying with time. Since the number of zero eigenvalues is equal to the number of connected components, if the number of \(\mathcal{L}\)'s zero eigenvalues increases, the diffusion ability of networks will get weakened. Figure 2 exhibits the proportion of none-zero eigenvalues of \(\mathcal{L}\) as a function of time steps. It is obvious that for the BA network, the number of none-zero eigenvalues increases till it reaches an upper bound, while for other networks such as the K-regular, the WS, and the ER, this value almost deceases to zero. Therefore, along with the diffusion process, more and more  nodes in the BA network get connected. On the contrary, in other networks, nodes are getting disconnected. This explains why the BA networks are more beneficial to the diffusion.


\section{Influence of network topology on diffusion}
\begin{figure}[!t]
  \captionsetup[subfigure]{labelformat=simple}
  \centering
  \includegraphics[width=2.5in]{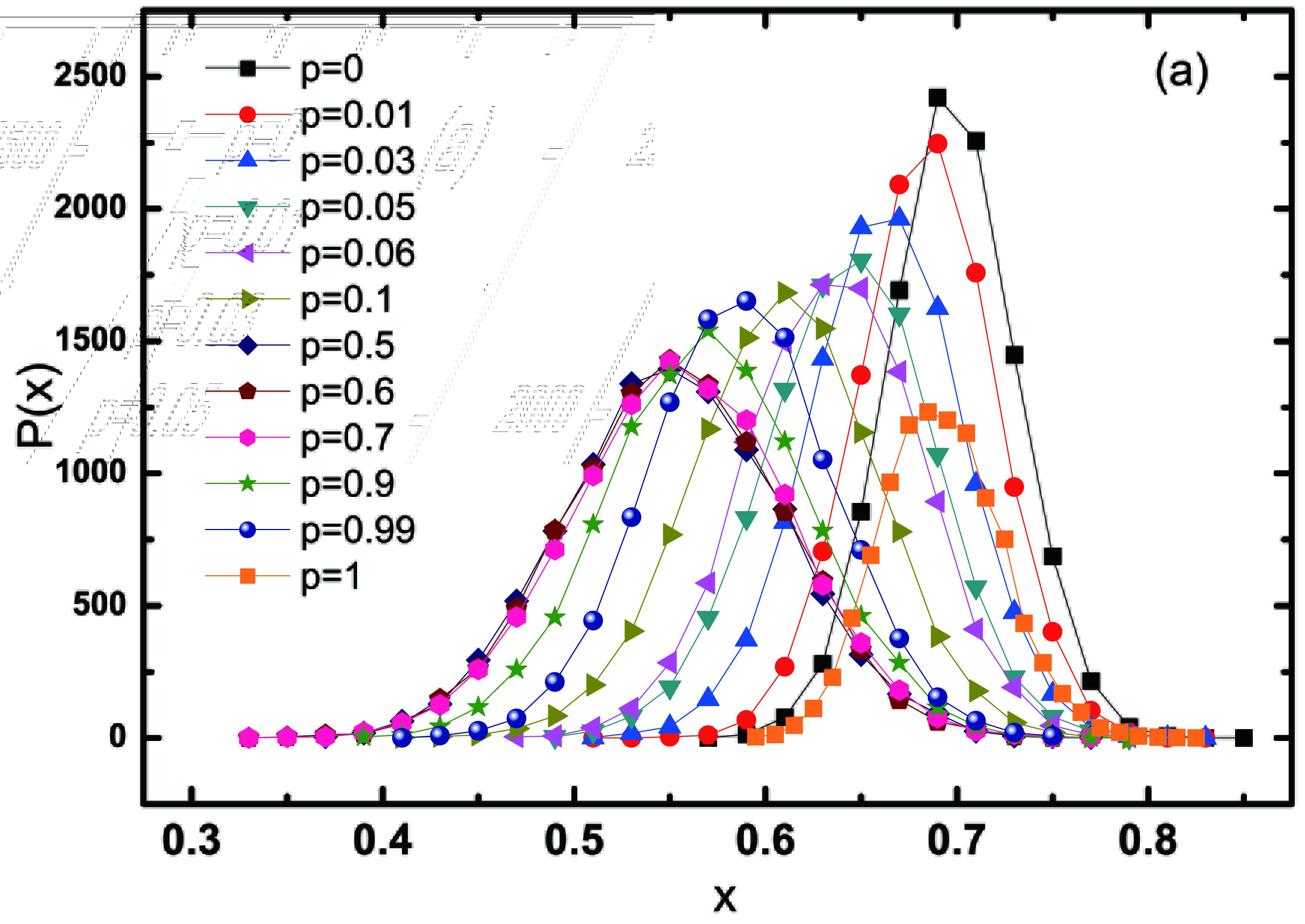}
  \includegraphics[width=2.5in]{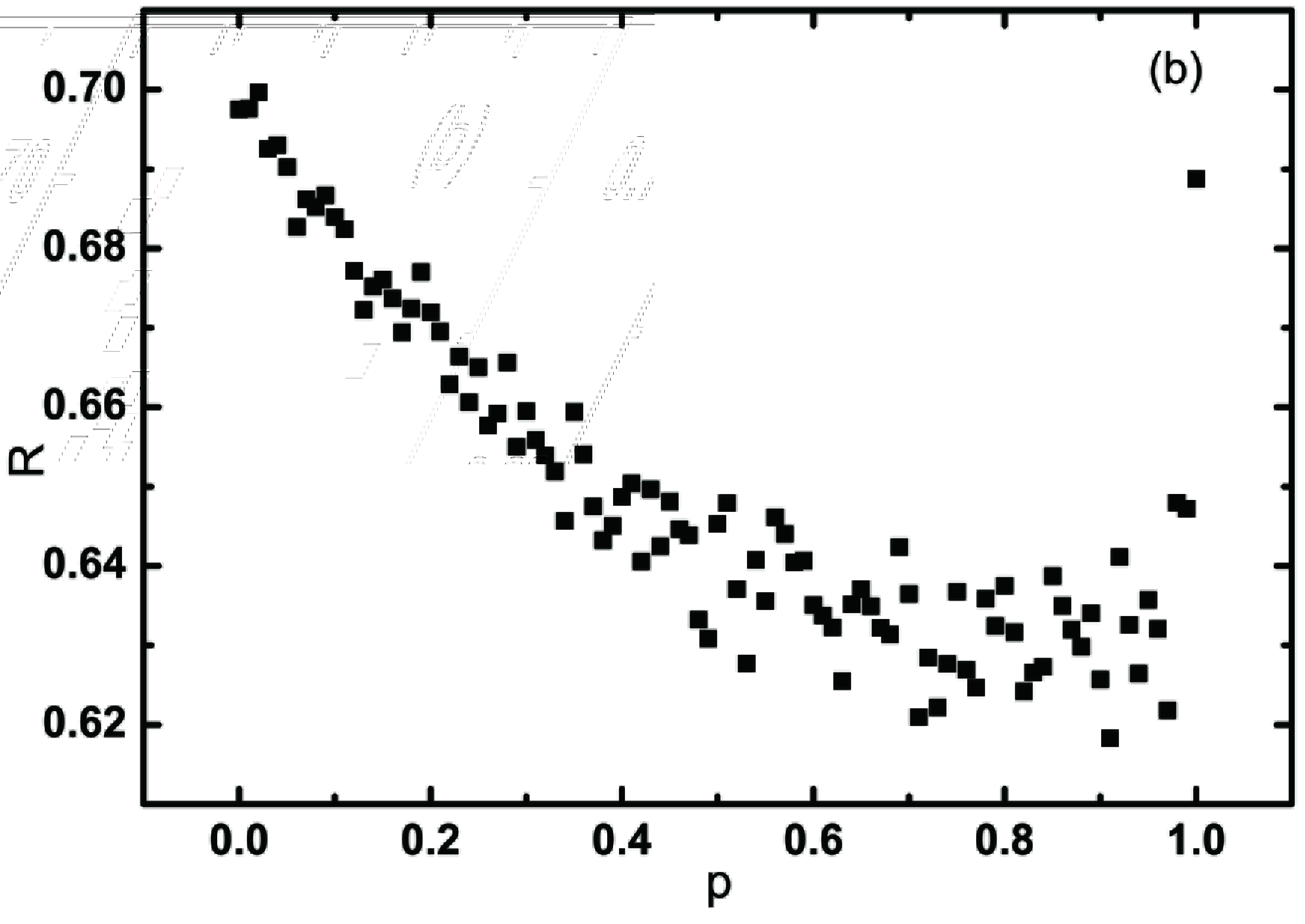}
  \caption{(a): The color dots and curves show the states' distribution \(P(x)\) as a function of state \(x\) with different rewiring probability \(p\); (b): The global state parameter \(R\) as a function of rewiring probability \(p\).}
  \label{fig:fig3}
\end{figure}

In this section, we study how the network topology affects the diffusion processes in Section II. First of all, we conduct an investigation into the state distributions of nodes after a diffusion process is stabilized under different network topologies. For heterogeneous networks such as the BA network, the states of nodes are distributed in a uniform value close to 1. In other words, almost every node in a BA network arrives at an identical state, which is approximately the scheduled upper bound. As for homogeneous networks, like the K-regular, the WS and the ER networks, the situations are different. Figure 3 displays the distribution of stabilized states for different choices of rewiring probability \(p\) in the Watts-Strogatz algorithm for generating WS small-world models~\cite{WS1998}. As is well known, when \(p\) changes, the networks can vary from K-regular (\(p=0\)), WS (\(p=0.05\)) to ER networks (\(p=1\))~\cite{Newman2010}. In addition, the states of nodes all exhibit a Gaussian-like distribution. Figure 3(b) plots the global state parameter \(R\) as a function of the rewiring probability \(p\), and shows that for K-regular and ER networks, the state values are relatively larger.

\begin{figure}[!t]
  \captionsetup[subfigure]{labelformat=simple}
  \centering
  \includegraphics[width=3.3in]{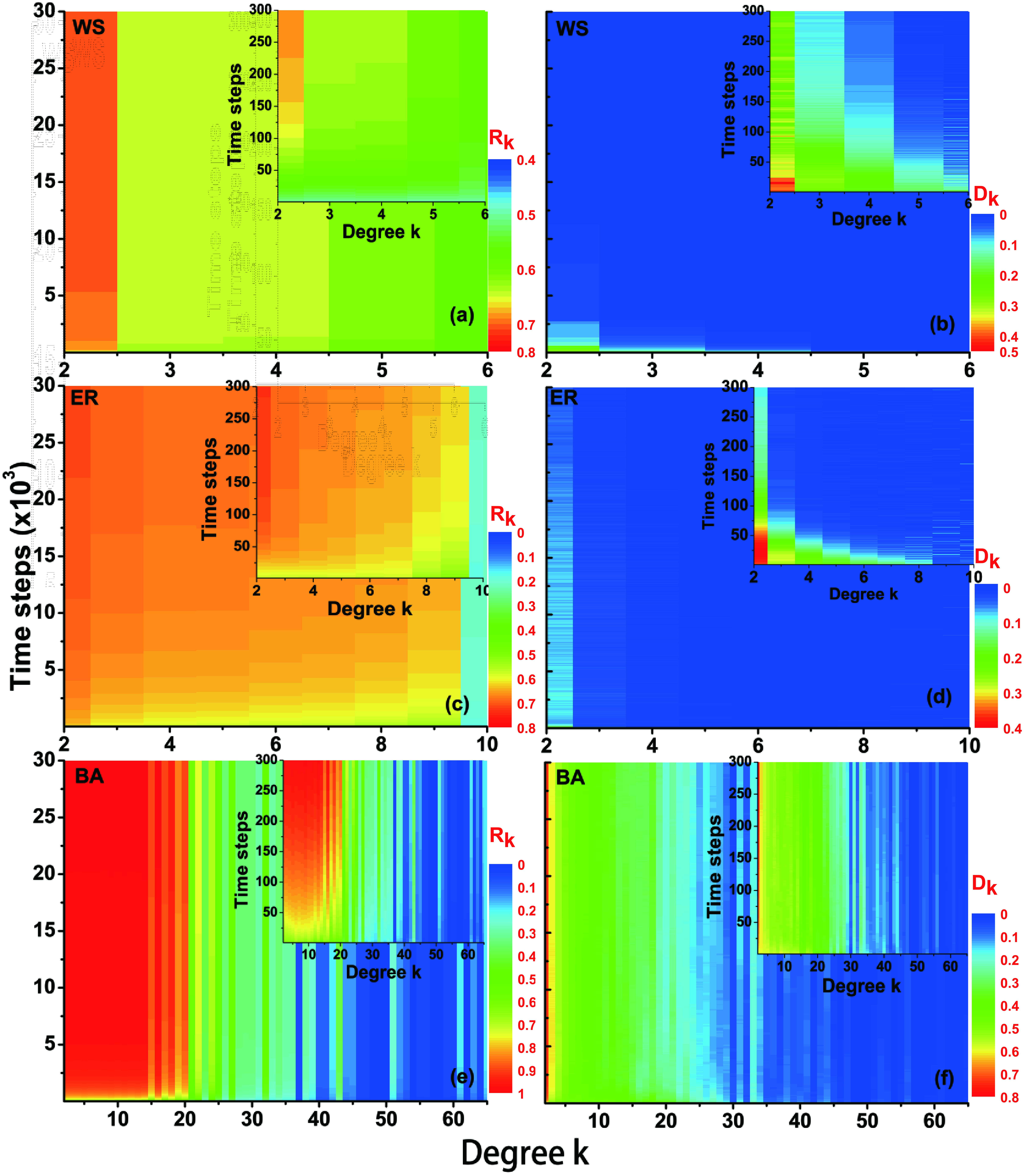}
  \caption{Degree-\(k\) nodes' average state \(R_{k}\) (left) and the diffusion coefficient \(D_{k}\)  (right) varying with time, for the WS (upper), the ER (middle) and the BA (bottom) networks.}
  \label{fig:fig4}
\end{figure}

From Fig. 3, we assume that the ability of a node to diffuse is related to its degree. Thus we reorganize the data and rank the nodes by their degrees, then get the average state parameter \(R_{k}\) and the average diffusion coefficient \(D_{k}\) for degree-\(k\) nodes. We record their changing processes during the diffusion and find some interesting results. Figure 4 shows the contour graphs of \(R_{k}\) and \(D_{k}\) as a function of \(k\) and \(t\) under different network topologies. The left panels of Fig. 4 display the degree-\(k\) nodes' average state \(R_{k}\) varying with time. We can obtain that small-degree nodes reach higher states than large-degree nodes do. Correspondingly, the small-degree nodes' diffusion ability is better than the large-degree nodes, and the nodes with large degrees tend to maintain their original states, as can be observed from the diffusion coefficient \(D_{k}\) in the right panels of Fig. 4. That means, small-degree nodes are susceptible to their neighbours and the large-degree nodes are much less susceptible.

Furthermore, there is difference between homogeneous and heterogeneous networks. For homogeneous networks, as can be seen from Panels (a) to (d), the proportion of small-degree nodes is relatively small, and most nodes have an average degree, with relatively weaker diffusion ability leading to a smaller global state value. While for heterogeneous networks, as can be obtained from Panels (e) and (f), the proportion of small-degree nodes is very large due to the power law degree distribution, and the nodes' strong diffusion ability leads to a large global state value.

\section{Minimization of state entropy during diffusion}
In fact, some diffusion processes, such as synchronization and consensus, are evolving processes from disorder to order. In order to gain a thorough understanding of the diffusion process we proposed, we introduce state entropy to measure the level of disorder of the dynamic system, which is defined as follows,

\begin{equation}\label{8}
S(Q)=-\sum^{N}_{s=1}q_{s}logq_{s},
\end{equation}
where \(Q=(q_{1}, q_{2}, \cdots, q_{N})^{T}\) and \(q_{s}\) (\(s=1,2, \cdots, N\)) is the probability that a node locates in \(s-\)th state, and \(N\) is the number of possible states. For a general linear diffusion process on a fully connected graph, \(N\) normally equals 1. Under this circumstance, state entropy reaches its minimal value, which is zero. While in this research, \(N\) equals to the number of zero eigenvalues of \(\mathcal{L}\) from Eq. (4), meaning that the number of possible states equals the number of connected components.

\begin{figure}[!t]
  \centering
 \includegraphics[width=3in]{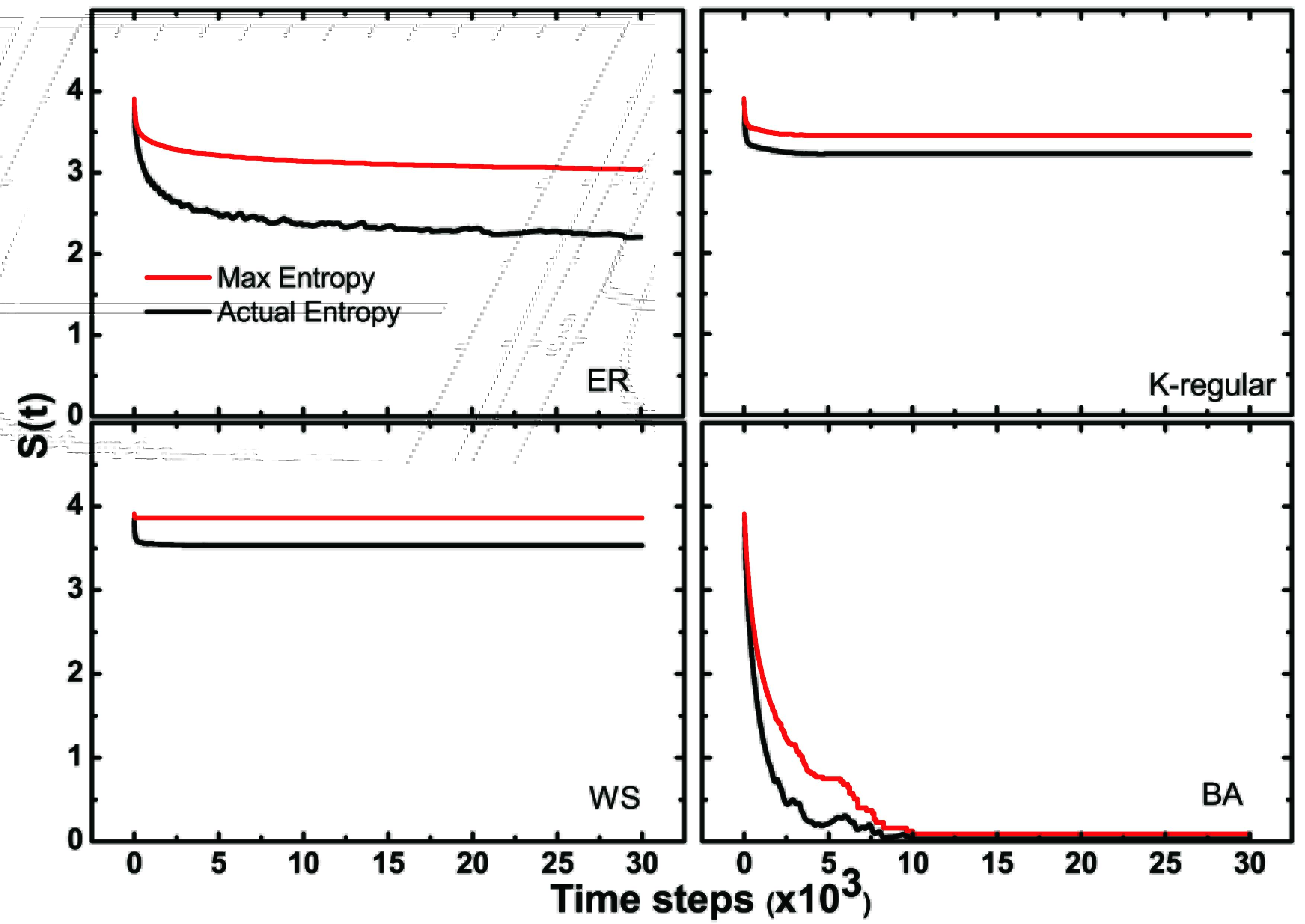}\\
     \caption{State entropy \(S(t)\) as a function of time steps under different network topologies. The red lines represent theoretical maximum entropy and black lines represent actual state entropy.}\label{entropy1}
\end{figure}

According to the maximum entropy principle, when \(q_{s}\) has a uniform distribution, e.g., \(q_{s}=1/N\), the system reaches its maximum entropy. Apparently, when \(N>1\), the distribution of \(q_{s}\) is unknown during the diffusion process. Now we assume that the diffusion we proposed is a process of minimization of state entropy. In Fig. 5, we plot state entropy described by Eq. (8) as a function of time for different network topologies, and compare with their theoretical maximum value according to the maximum entropy principle. The plot exhibits a minimization process of state entropy during the diffusion. Similarly, in Fig. 6, when diffusion is stabilized, we plot the state entropy as a function of rewiring probability \(p\), and obtain similar observations. For the BA network, its state entropy can reach a much smaller value compared with other networks, it is consistent with the result that BA network is beneficial to diffusion we proposed.

\begin{figure}[!t]
  \centering
 \includegraphics[width=2.5in]{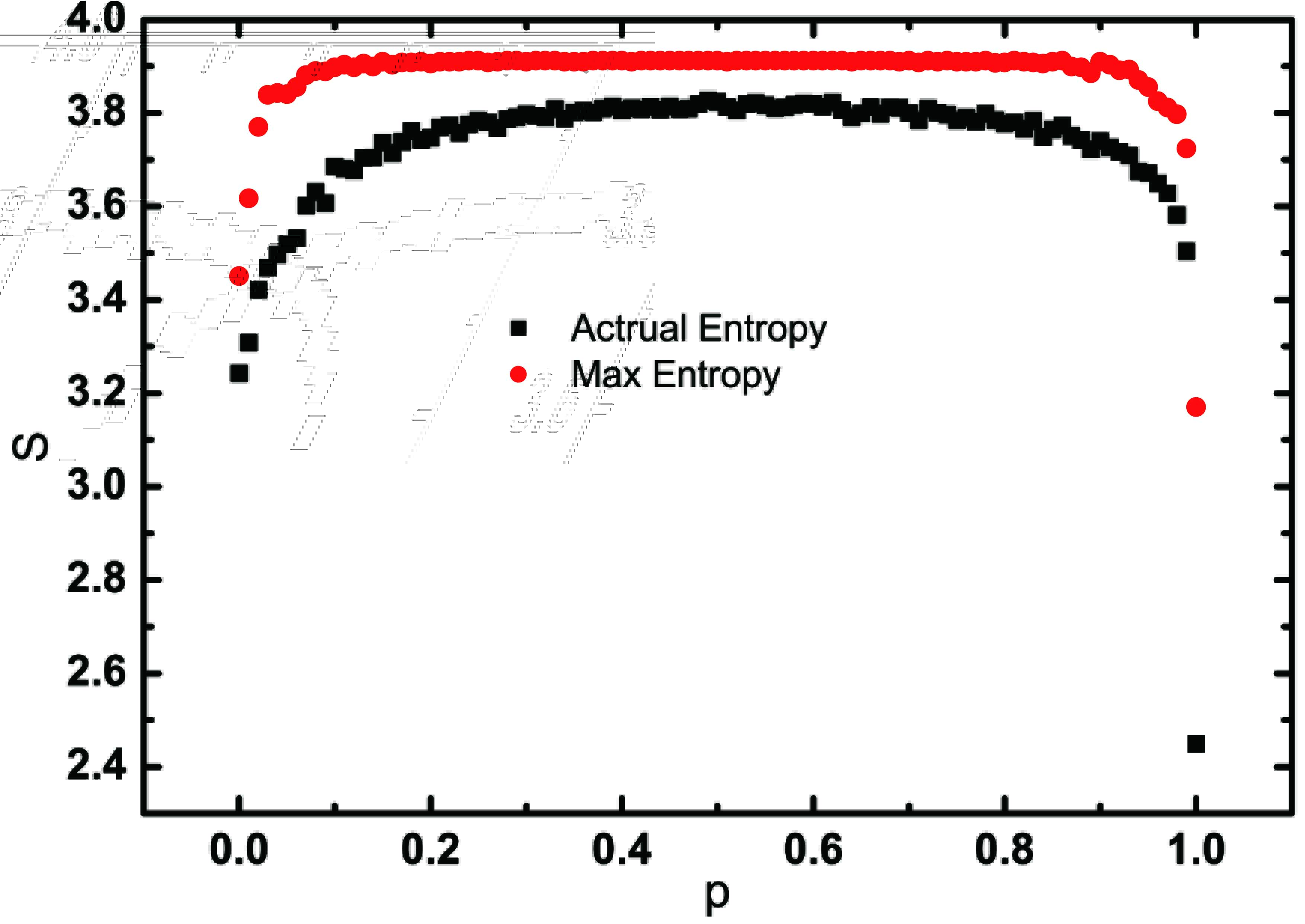}\\
     \caption{Stabilized state entropy \(S\) at \(t=30000\) as a function of rewiring probability \(p\). The red dots represent the theoretical maximum entropy and black dots represent actual state entropy.}\label{entropy2}
\end{figure}

Now consider the minimization of state entropy as an inequality constrained optimization problem, then we have

\begin{equation}\label{9}
\begin{cases}
\mbox{min}S(Q)=-\sum^{N}_{s=1}q_{s}logq_{s},\\
\mbox{s.t.}\ g(Q)=\sum^{N}_{s=1}q_{s}=1,\\
\mbox{s.t.}\ h(Q)\leq0,
\end{cases}
\end{equation}
where \(g(Q)\) and \(h(Q)\) are constraint functions. In order to obtain the minimum value of \(S(Q)\), we use the Lagrange multiplier method and have

\begin{equation}\label{10}
\nabla S(Q)+\nabla \lambda (g(Q)-1)+\nabla \lambda^{\prime}h(Q)=0,
\end{equation}
where \(\nabla\) is the vector differential operator. Eq. (10) meets the KKT condition,

\begin{equation}\label{11}
\begin{cases}
g(Q^{*})-1=0,\\
h(Q^{*})\leq0,\\
\lambda^{\prime}\geq0,\\
\lambda^{\prime}h(Q^{*})=0,
\end{cases}
\end{equation}
where \(Q^{*}\) is the feasible point and function \(h(Q)\) is unknown.

Next, taking the partial derivatives with respect to each \(q_{s}\), we have for \(s\in(1,...,N)\),

\begin{equation}\label{12}
-(\frac{1}{ln10}+logq_{s})+\lambda+\frac{d}{dq_{s}}\lambda^{\prime}h(q_{s})=0.
\end{equation}

Eq. (12) can be written as

\begin{equation}\label{13}
dh(q_{s})=[\frac{\lambda}{\lambda^{\prime}}(\frac{1}{ln10}+logq_{s})-\frac{\lambda}{\lambda^{\prime}}]dq_{s}.
\end{equation}

Integrating Eq. (13), then we have

\begin{equation}\label{14}
h(q_{s})=\alpha q_{s} +\beta q_{s}logq_{s} +\gamma,
\end{equation}
where \(\alpha=\frac{1}{\lambda^{\prime}ln10}+\frac{\lambda-1}{\lambda^{\prime}}\), \(\beta=\frac{1}{\lambda^{\prime}}\) and \(\gamma=\frac{C^{\prime}}{\lambda^{\prime}}\) (\(C^{\prime}\) is a constant value).

Then Eq. (14) can be written as

\begin{equation}\label{15}
h(Q)=\sum^{N}_{s=1}(\alpha q_{s} +\beta q_{s}logq_{s} +\gamma)=\beta\sum^{N}_{s=1}q_{s}logq_{s}+C.
\end{equation}

Finally, we have

\begin{equation}\label{16}
h(Q)=C-\beta S(Q),
\end{equation}
according to the KKT condition, where \(\beta>0\) and \(C>0\).

We can obtain that Eq. (16) has the same form as the Gibbs free energy, \(G=H-T\cdot S\), where \(S\) is the entropy, \(T\) is the absolute temperature and \(H\) is the enthalpy. According to the thermodynamics, during the entropy minimization process restricted by Gibbs free energy, when \(\bigtriangleup h(Q)>0\), \(\bigtriangleup S(Q)<0\) and \(C>0\), the diffusion process is a non-spontaneous process. This finding opens up a new way to investigate thermodynamic representations in networks and provides clear directions for optimizing relevant dynamical processes on complex systems.

\section{Conclusion}
In this paper, we have introduced a diffusion model adapted to a network's local topological and dynamical information, which varies with time. We have studied the diffusion process under different network topologies and found that heterogeneous networks, such as the BA network, are beneficial to the diffusion process and lead state values to reach a higher level than homogeneous networks do. Furthermore, we have investigated the influence of network topologies on the diffusion, and found that the distribution of stabilized states are related to the network's degree distribution. During this kind of diffusion process, the nodes with smaller degrees are more likely to change and reach a larger state values, while those who have larger degrees tend to hold their intrinsic states. To further understand the thermodynamic mechanism of the diffusion process, we have introduced a concept of state entropy and found that minimization of state entropy exists during the diffusion. Finally, by using the inequality constrained optimization method, we have obtained a restriction function bearing the same form as the Gibbs free energy, and pointed out that the diffusion process we proposed is a non-spontaneous process. The findings encourage us to study and understand dynamics on complex networks from different perspectives and could be used to uncover hidden statistical features of complex systems.


%

\section*{Acknowledgment}
This work was supported by the National Natural Science Foundation of China under Grants 61573262, and in part by the Natural Science Fund for Distinguished Young Scholars of Hubei Province under Grant 2017CFA052.

\ifCLASSOPTIONcaptionsoff
  \newpage
\fi


\begin{thebibliography}{1}

\bibitem{Barabasi1999}Barabasi, A. L., \& Albert, R.
Emergence of scaling in random networks.
{\it science} {\bf 286}, 509--512 (1999).

\bibitem{WS1998}Watts, D. J., \& Strogatz, S. H.
Collective dynamics of `small-world' networks.
{\it Nature} {\bf 393}, 440--442 (1998).

\bibitem{ER1960}Erd\"{o}s, P., \& R\'{e}nyi, A.,
On the evolution of random graphs.
{\it Publ. Math. Inst. Hungar. Acad. Sci.} {\bf 5}, 17-61 (1960).

\bibitem{Newman2010}Newman, M. E. J.
Networks: An Introduction.
{\it OxfordUniversity Press, Oxford, (2010)}.

\bibitem{Masuda2017}Masuda, N., Porter, M. A., \& Lambiotte, R.
Random walks and diffusion on networks.
{\it Phys. Rep.} (2017).

\bibitem{Pastorsatorras2015}Pastorsatorras, R., Castellano, C., Mieghem, P. V., \& Vespignani, A.
Epidemic processes in complex networks.
{\it Review of Modern Physics} {\bf 87(3)}, 120-131 (2015).

\bibitem{Arenas2008} Arenas, A., D\'iaz-Guilera, A., Kurths, J., Moreno, Y. \& Zhou, C.
Synchronization in complex networks.
{\it Phys. Rep.} {\bf 469}, 93--153 (2008).

\bibitem{Barabasi2016} Barab\'{a}si A.-L.
Network Science
{\it(Cambridge University Press, Cambridge, 2016)}.

\bibitem{Pearson1905}Pearson, K.
The problem of the random walk.
{\it Nature} {\bf 72(1865)}, 294 (1905).

\bibitem{Porter2014}Porter, M. A., \& Gleeson, J. P.
Dynamical systems on networks: a tutorial.
{\it Physics} (2014).

\bibitem{Kermack1927}Kermack, W., McKendrick, A.
A contribution to the mathematical theory of epidemics.
{\it P R Soc Lond A-Conta} {\bf 115}, 700¨C721 (1927).

\bibitem{Pastorsatorras2001}Pastorsatorras, R., \& Vespignani, A.
Epidemic spreading in scale-free networks.
{\it Phys. Rev. Lett.} {\bf 86(14)}, 3200-3203 (2001).

\bibitem{Wei2018}Wei, X., Wu, X., Chen, S., Lu, J., \& Chen, G.
Cooperative epidemic spreading on a two-layered interconnected network.
{\it SIAM Journal on Applied Dynamical Systems}, {\bf 17} ,1503-1520 (2018).

\bibitem{Liyang2015}Li, Y., Wu, X., Lu, J. A. \& L\"{u}, J.
Synchronizability of duplex networks.
{\it IEEE Trans. Circuits Sys. II: Express Briefs} {\bf 63}, 206-210 (2015).

\bibitem{Tang2016}Tang, L., Wu, X., L\"{u}, J., Lu, J., \& D'Souza, R. M.
Master stability functions for multiplex networks.
arXiv:1611.09110v1 (2016).

\bibitem{Mei2017}Mei, G., Wu, X., Wang, Y., Hu, M., Lu, J. A., \& Chen, G.
Compressive-sensing-based structure identification for multilayer networks.
{\it IEEE Transactions on Cybernetics} {\bf 48(2)}, 754-764 (2017).

\bibitem{Zhuang2018}Zhuang, J., Cao, J., Tang, L., Xia, Y., \& Perc, m.
Synchronization analysis for stochastic delayed multilayer network with additive couplings.
{\it IEEE Transactions on Systems, Man, and Cybernetics: Systems}, 1-10 (2018).

\bibitem{Almendral2007}Almendral, J. A., \& D¨ªazguilera, A.
Dynamical and spectral properties of complex networks.
{\it Physics} {\bf 9(26)}, (2007).

\bibitem{Daido1987}Daido, H.
Population dynamics of randomly interacting self-oscillators. I: Tractable models without frustration.
{\it Progress of Theoretical Physics} {\bf 77(3)}, 622-634 (1987).

\bibitem{Lumer1991}Lumer, E. D., \& Huberman, B. A.
Hierarchical dynamics in large assemblies of interacting oscillators.
{\it Physics Letters A} {\bf 160(3)}, 227¨C232 (1991).

\bibitem{Filatrella2007}Filatrella, G., Pedersen, N. F., \& Wiesenfeld, K.
Generalized coupling in the kuramoto model.
{\it Phys. Rev. E} {\bf 75(2)}, 017201 (2007).

\bibitem{Barahona2002}Barahona, M., \& Pecora, L. M.
Synchronization in small-world systems.
{\it Phys. Rev. Lett.} {\bf 89(5)}, 054101 (2002).

\bibitem{Park2004}Park, J., \& Newman, M. E.
Statistical mechanics of networks.
{\it Phys. Rev. E} {\bf 70(2)}, 066117 (2004).

\bibitem{Dehmer2011}Dehmer, M., \& Mowshowitz, A.
A history of graph entropy measures.
{\it Information Sciences} {\bf 181(1)}, 57-78 (2011).


\end{thebibliography}
\end{document}